\documentclass[nologo,11pt,a4paper,hidelinks]{ETHpaper}

\usepackage{tabularx}
\usepackage{MnSymbol}
\usepackage{booktabs}
\usepackage{caption}
\usepackage{hhline}
\usepackage[table,dvipsnames]{xcolor}
\usepackage{graphicx}
\usepackage[english]{babel}
\usepackage[numbers,sort&compress]{natbib} 
\usepackage{setspace}
\usepackage{textcmds}
\usepackage{amsmath}

\title{Group relations, resilience and the \emph{I Ching}}
\titlealternative{Group relations, resilience and the \emph{I Ching}} 
\author{Frank Schweitzer}
\authoralternative{Frank Schweitzer}
\address{Chair of Systems Design, ETH Zurich, Switzerland\\
 Complexity Science Hub Vienna, Austria} \www{\url{http://www.sg.ethz.ch}}

\reference{Physica A (accepted for publication, 2022) } 
\makeframing

\usepackage{appendix}
\usepackage[utf8]{inputenc}
\usepackage{xcolor}
\usepackage{soul}
\usepackage[hang]{subfigure}
\usepackage{multirow}
\usepackage{hyperref}
\usepackage{xparse,tikz,calc} \usepackage{pifont}

\renewcommand{\epsilon}{\varepsilon}

\newcommand{\abs}[1]{\left| #1 \right|}
 \newcommand{\sign}{\operatorname{sign}}

\begin{document}

  \maketitle

\begin{abstract}
  We evaluate the robustness and adaptivity of social groups with heterogeneous agents that are characterized by their binary state, their ability to change this state, their status and their preferred relations to other agents.   
  To define group structures, we operationalize the hexagrams of the \emph{I Ching}.
  The relations and properties of agents are used to quantify their influence according to the social impact theory.
  From these influence values we derive  a weighted stability measure for triads involving three agents, which is based on the weighted balance theory. 
  It allows to quantify the robustness of groups and to propose a novel measure for  group resilience which combines robustness and adaptivity.
  A stochastic approach determines the probabilities to find robust and adaptive groups.
  The discussion focuses on the generalization of our approach. 
\end{abstract}

\section{Introduction}
\label{sec:introduction}

The aim of our paper is a formal analysis of the relations and the resilience of \emph{social groups}.
Different from communities, 
these groups are of rather small size which requires to consider the properties of their members, denoted as agents in the following, in more detail.
In social systems these agents are \emph{heterogeneous}, i.e., they have a different character, social status, influence on others.
We are particularly interested in the feedback between agent properties and their relations because of its social relevance. 
For instance, studies about political polarization acknowledge a feedback between the opinions of individuals and the evolution of their mutual relations which may then lead to an increasing polarization \citep{Lee2016,Schweitzer_2020,Starnini2021}.
But a framework that combines the internal dynamics with the evolution of relations is still under development \citep{Schweighofer_2020,Gao2018,Deng2016,Chen2014,Singh2016,Flache2015}.  

To model the impact of heterogeneous properties on other agents and the group as a whole, we need  assumptions about the group structure, its dynamics of change, the relations between agents and their importance.
Instead of arbitrary ad hoc assumptions we resort on the \emph{I Ching}, a classic Chinese text about \qq{change} introduced in Section~\ref{sec:i-ching}. 
Specifically, we interpret the hexagrams of the \emph{I Ching} as descriptions of group relations.
We note that our analysis can be easily generalized if other information about agents and their relations shall be taken into account.

Our work contributes to a formal modeling of \emph{groups} as the smaller building blocks of social systems.
While large social networks can be reasonably described by a statistical approach that averages over many individual details, the formation of groups requires to understand the social mechanisms that lead individuals to join or to leave, to form or to delete links \citep{Schweitzer2014,Groeber2009,uzzi2019,Gorski2017,Gao2018}. 
Social network analysis has already studied some of these mechanisms, for instance reciprocity, homophily and triadic closure among individuals \citep{Gorski2019,Harrigan2017,Deng2016,Yap2015}. 
Recent work also discusses the costs and benefits of being part of a group to derive  individual utility functions \citep{nucleation21}.
These factors are important to assess the stability of groups as a precondition to their further growth and merger into larger networks.
At the same time, we also need to consider the impact of relations on the stability of a social system.

The core of our approach is a new measure for the stability and the resilience of groups. 
In line with other works \citep{antal2,antal3,Gao2018,Belaza2017,Gorski2019} we estimate group stability from the stability of so called triads, i.e., building blocks of groups involving three agents. 
Triadic closure is seen as a generative mechanism for social communities \citep{Gao2018,Gorski2019,brandenberger2019,uzzi2019,Kulakowski2019,Sherwin1975}. 
Models for structural balance \citep{Agbanusi2018,antal2,uzzi2019,aref2017measuring,Belaza2017} determine the stability of these triads from the signs of the relations between agents  (see Figure~\ref{fig:triads}).
To ease analytic investigations,  these triads are often considered to be independent \citep{antal3,Belaza2017}, ignoring the fact that agents are part of different triads at the same time.

Different from these works, to calculate the group stability we first weight these triads by considering the importance and the social impact of the involved agents.
For this, we build on two social theories.
The \emph{social impact theory} \citep{Bushman2007,Jackson1987,fink-96,lewenst-nowak-latane-92,nowak-szam-latane-90,Jackson1987,Finsterbusch1982,latane-81} tries to quantify the influence that a focal agent experiences from other agents in the group.
This depends on the status of the agents involved (strength), their direct or indirect relations (immediacy) and their supporting or opposing attitude. 
Hence, with our computation of \emph{influence values} for each agent we already have their mutual relations and their status taken into account.
To subsequently determine the weight of a triad, we use the \emph{weighted balance theory} \citep{Schweighofer_2020}.
It extends Heider’s cognitive balance theory \citep{heider1946attitudes} to encompass multiple weighted attitudes.
In our context this approach allows to weight the influence values of agents in addition to the signs of their relations.
In other words, agents contribute to a different degree to the in/stability of triads, which can now be quantified with a continuous measure instead of a dichotomous distinction between stability and instability.

This fine grained stability measure allows us to further quantify the resilience of a group.
Resilience differs from conventional notions of stability because it additionally reflects the ability to respond to change.
To capture this function, we introduce two dimensions of resilience, namely robustness and adaptivity.
Robustness, as the structural component of resilience, can be derived from our stability measure.
Adaptivity, the dynamic component of resilience, is a two-edged sword.
It bears the chance to improve the robustness of the group, but also the risk to reduce it.
If a robust system changes too much it can lose its robustness.
On the other hand, the ability to adapt is beneficial for systems with low robustness, because it allows them to obtain a better state. 
Hence, different from stability, resilience also considers the ability to improve in a \emph{future state}.

The paper is organized as follows.
In Section~\ref{sec:small-groups}, we introduce the basic dynamics of our agent-based model and discuss the structural balance of triads. 
Agents are characterized by a binary internal state which can change under certain circumstances.
Relations between agents may depend on their states which indirectly determine the stability of triads.  
A didactic example presented in the Appendix further illustrates the concept and the resulting problems.
In Section~\ref{sec:i-ching}, we summarize facts about the \emph{I Ching} and explain the hexagrams composed of 6 lines, which we take as descriptions of group structures.
To map explanations from the \emph{I Ching} with our agent-based model, we formalize 
the concept of the \emph{I Ching}. 
This  is a challenge on its own, described in Section~\ref{sec:i-ching}, which will lead us to a description of the signed relations in a small group of 6 agents. 

In Section~\ref{sec:eval-group-struct}, we build on these relations to calculate the social impact for each agent based on the supporting and opposing influences from other agents.
Taking the status into account, this determines the influence of each agent.
In Section~\ref{sec:weight-balance-cond} we use ideas from weighted balance theory to calculate the weighted stability of triads and the resulting stability of the group. 
This allows us in Section~\ref{sec:resil-small-groups} to estimate the group resilience.
A stochastic approach presented in Section~\ref{sec:stochastic-approach} informs about the probability to find certain group configurations and about their possible change.
Section~\ref{sec:disc} is devoted to a detailed discussion of our modeling approach and possible extensions.

\section{Dynamics in small groups}
\label{sec:small-groups}

Let us consider a small group of $N=6$ agents which are identified by their number
$i\in\{1,2, ...,6\}$.  Each agent can have relations to other agents, which are represented as a
network (see Figure \ref{fig:GG}).  Agents are the nodes in this network and the variable
$w_{ij}\in \{-1,0,+1\}$ indicates their relation.
Positive relations, for instance support or consensus, are indicated by $w_{ij}=+1$, negative relations, for instance conflict or repression, by $w_{ij}=-1$.
$w_{ij}=0$ if agents in the network are not connected.
Each agent is characterized by a binary state variable $s_{i}\in \{-1,+1\}$, which may change over time. 
Whether this possibility exists is indicated by a variable $a_{i}\in \{-1,+1\}$, which denotes the ability to change. 
In this paper, we consider only two times, $t_{0}$ before and $t_{1}$ after the change. 
The dynamics for each agent reads
\begin{align}
  \label{eq:1}
  s_{i}(t_{1})=s_{i}(t_{0})a_{i}
\end{align}
With $a_{i}=-1$, the state at $t_{1}$ becomes the opposite of the state at $t_{0}$, with $a_{i}=+1$,
the state does not change.
We note that the value $a_{i}=-1$ only occurs with a rather small probability which is determined in Section~\ref{sec:basic-features}. 
The group configuration  is expressed by a vector
$ \mathbf{S}(t)=\{s_{1},s_{2},...,s_{6}\}$. 
We can then compare the current situation of the small group, \emph{before} the change, given by $\mathbf{S}(t_{0})$, with the situation in the near future, \emph{after} the change, given by $\mathbf{S}(t_{1})$.

Before we can discuss the stability of the group and how it is affected by the change, we have to determine the relations $w_{ij}$ and their dependence on the  agent variables.
This is carried out in detail in Section~\ref{sec:i-ching}. 
In the Appendix, we further discuss a didactic example of a fully connected network where the states $s_{i}$ are interpreted as agents' opinions.
The relations result from $w_{ij}=s_{i}s_{j}$.
I.e,. they are positive if agents have the same opinion, denoted as consensus, and negative if they have different opinions.
Assuming that consensus positively affects stability, 
such a setup allows to investigate whether change would lead to more or to less stable groups.
This can be captured by measuring the fraction of positive relations in the network at the two different time steps.

\begin{figure}[htbp]
  \centering \includegraphics[width=0.99\textwidth]{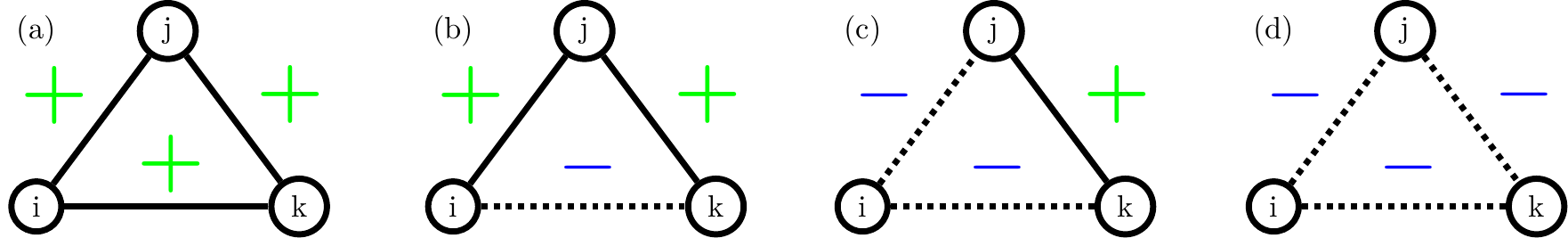}
  \caption{Possible configurations of the $w_{ij}$ in triads $T_{ijk}$.  (a), (c) are considered
    stable triads, (b), (d) unstable triads following Eqn.~\eqref{eq:9}.}
  \label{fig:triads}
\end{figure}

Models of structural balance often assign the values of the binary relations $w_{ij}\in\{+1,-1\}$
\emph{randomly}, i.e. without considering the states of agents $s_{i}$ \citep{Belaza2017,antal2,antal3}.  The focus is then on the
emerging triads $T_{ijk}$, involving any three agents $i,j,k$, as shown in Figure~\ref{fig:triads}.
Whether a triad $T_{ijk}$ is assumed as structurally stable or unstable depends on the product of the signs of 
the relations:
\begin{align}
  \label{eq:9}
  T_{ijk}=w_{ij}\,w_{ik}\,w_{kj}=
  \begin{cases}
    +1 & \mathrm{stable} \\
    -1 & \mathrm{unstable} \\
  \end{cases}
\end{align}

Most models of structural balance postulate a dynamics in which agents have the ability to change the signs of 
their $w_{ij}$ such that an unstable triad becomes a stable one \citep{Agbanusi2018,Du2018,Gao2018,Belaza2017,antal2,antal3}.
To capture the change, one can compare the fraction of balanced triads in the network as shown in the Appendix. 
We want to deviate from this approach by considering a feedback between agents' states $s_{i}$ and their relations $w_{ij}$ in a more subtle manner.
This is where the \emph{I Ching} comes into play.
Precisely, we will interpret the hexagrams from the \emph{I Ching} as group structures, as discussed in Section~\ref{sec:relat-betw-agents}.

\section{The \emph{I Ching}}
\label{sec:i-ching}

\subsection{Basic features}
\label{sec:basic-features}

\begin{figure}[htpb]
  \centering \includegraphics[width=0.7\textwidth]{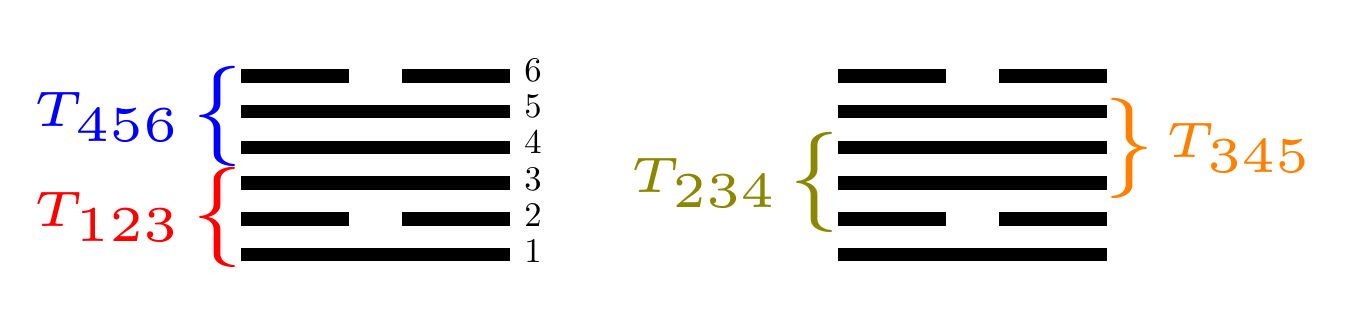}
  \caption{Hexagram with primary trigrams $T_{123}$, $T_{456}$ and nuclear trigrams $T_{234}$,
    $T_{345}$.}
  \label{fig:hex}
\end{figure}

The \emph{I Ching}, or \emph{Yi Ying}, translated as \qq{The Book of Change}, is a classic Chinese
text with roots from more than 2000 years ago.  It tries to describe a current situation by means of
\emph{hexagrams}, i.e. symbols composed of \emph{six} lines (see Figure~\ref{fig:hex}).  It is not
the purpose of this paper to summarize the origins, developments and multifaced interpretations of
the \emph{I Ching}.  We restrict ourselves to the very basics needed to explain its relation to
group formation, the way we see it.  The interested reader is referred to the extensively commented
\emph{I Ching} edition by Richard Wilhelm \citep{WB}, from which we quote in the following.  It was translated
to English by Cary Baynes and still serves as a main reference point, despite newer editions,
because of its deep and insightful explanations.

The six lines in each hexagram form a hierarchy from bottom to top, indicated by their ascending
\emph{places} 1,...,6.  Each line can appear in two different states, indicated by a broken and an
unbroken line.  Broken lines are also referred to as weak, earthly, receptive, dark, negative,
yielding, \emph{yin}.  Unbroken lines are referred to as strong, heavenly, easy, creative, light,
positive, firm, \emph{yang}.  One has to refrain from associating these lines with notions of
\qq{good} or \qq{bad}.  According to Chinese philosophy, the principles of \emph{yin} and
\emph{yang} are \emph{complementary} rather than contradictory, i.e. both are needed to constitute
the world and their balance is essential for harmony.  Subtle relations to the Greek concepts of
\emph{logos} and \emph{eros} cannot be denied.

\begin{figure}[htpb]
  \centering
  \scalebox{4.5}{\includegraphics{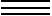}} { (a)} \hfill
  \scalebox{4.5}{\includegraphics{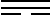}} { (b)} \hfill \scalebox{4.5}{\includegraphics{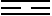}} { (c)} \hfill
\scalebox{4.5}{\includegraphics{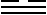}} { (d)} \\[1cm]

\scalebox{4.5}{\includegraphics{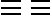}} { (e)} \hfill
\scalebox{4.5}{\includegraphics{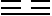}} { (f)} \hfill
\scalebox{4.5}{\includegraphics{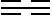}} { (g)} \hfill \scalebox{4.5}{\includegraphics{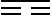}} { (h)}
  \caption{8 possible trigrams: (a)-(d) dark, (e)-(h) light trigrams.}
  \label{fig:tri2}
\end{figure}

From the combinations of broken and unbroken lines $2^{6}=64$ possible hexagrams result.  They are
seen as combinations of lower order structures, \emph{trigrams} of three lines each.  The $2^{3}=8$
possible trigrams are shown in Figure~\ref{fig:tri2}.  They can be divided into two groups.  The
\emph{dark} trigrams, shown in the top row, contain more solid than broken lines, while the
\emph{light} trigrams from the bottom row contain more broken than solid lines.

As shown in Figure~\ref{fig:hex}, each hexagram is formed by a \emph{lower} (inner) trigram,
$T_{123}$, and an \emph{upper} (outer) trigram, $T_{456}$.  These are referred to as \emph{primary}
trigrams.  Thus, the 64 hexagrams can be conveniently placed in a $8\times 8$ matrix, which has both
the rows and columns ordered by the 8 trigrams.  The correct order for the hexagrams is one of the
most debated issues in the \emph{I Ching} literature.  As a side note, Gottfried Wilhelm Leibniz,
who developed early concepts of binary numbers, became interested in the \emph{I Ching} after the
French Jesuit Joachim Bouvet sent him a map with the 64 hexagrams in \emph{Fu Xi} order, which
reflects their binary values.  Following the convention, we use instead the so called \emph{King
  Wen} order.  It assigns to each hexagram a number from 1 to 64, which is used for reference.
Additionally, each hexagram has a \qq{name}, for example No. 21 \emph{Shih Ho} \qq{Biting Through},
a \qq{judgement} and an \qq{image}.

It is important to note that the meaning of a hexagram cannot be simply decomposed into the meaning
of the constituting two trigrams, even less into the meaning of the constituting six lines.  As
shown in Figure~\ref{fig:hex}, each hexagram also contains two inner trigrams, referred to as
\emph{nuclear} trigrams, $T_{234}$ and $T_{345}$.  Therefore, the structure of each hexagram
reflects the overlapping influence of four trigrams in total, which we make use of in
Section~\ref{sec:relat-betw-agents}.

The name \qq{Book of Change} comes from the fact that the lines, under certain conditions, can
change their character, from \emph{yin} to \emph{yang}, or from \emph{yang} to \emph{yin}.  To
explain these conditions we note that the assignment of lines to places occurs by means of a random
process.  The abridged procedure uses 3 coins.  The \emph{heads} of each coin, which usually shows
the portrait of the ruler, counts 2, the \emph{tails} which usually shows the value, counts 3.  The
character of each line is determined by flipping these 3 coins together in one toss.  Summing up
the outcome for the three coins can only result in numbers 6, 7, 8, or 9, albeit with different
probabilities.  Table~\ref{tab:6-9} shows the possible combinations.  The \emph{even} numbers 6 and
8 refer to \emph{yin} and are represented by a broken line, the \emph{odd} numbers 7 and 9 refer to
\emph{yang} and are represented by a non-broken line.  6 and 9 appear with a smaller probability.
They are referred to as the \qq{old \emph{yin}} and the \qq{old \emph{yang}}, which can change their
character into the (young) \emph{yang} and the (young) \emph{yin}, respectively.

\begin{figure}[htbp]
  \centering {\scalebox{4.5}{\includegraphics{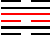}}}(a) \hspace*{0.5cm}
  \raisebox{0.5cm}{\fontsize{45}{5}{$\Rightarrow$}} \hspace*{0.5cm}
  {\scalebox{4.5}{\includegraphics{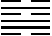}}}(b)
  \caption{(a) Initial hexagram No. 47, red lines are changing (b) resulting hexagram No. 48.}
  \label{fig:change}
\end{figure}

A hexagram is built \qq{bottom up} by consecutively assigning lines to places in ascending order.
Hexagrams can only change if they contain lines representing the numbers 6 or 9.  We have decided to
color such lines in red, for convenience.  Figure~\ref{fig:change} illustrates the change using the
hexagram No. 47 \emph{K’un} \qq{Oppression (Exhaustion)}.  In the \emph{I Ching}, a commentary about
lines and places is \emph{only} given if a respective place contains a line that can \emph{change}.
All other lines, while being important to constitute the hexagram, do not receive an interpretation.

For illustrative purposes, this hexagram has by chance  \qq{Six in the third place} and \qq{Nine in
  the fourth place}.  The commentary then states: \qq{Six in the third place means: A man permits
  himself to be oppressed by stone, And leans on thorns and thistles.  He enters his house and does
  not see his wife.  Misfortune} \citep[p. 407]{WB}. \qq{Nine in the fourth place means: He comes very quietly,
  oppressed in a golden carriage.  Humiliation, but the end is reached} \citep[p. 408]{WB}.
Obviously,
discovering the meaning of these comments is beyond our interest.  They are
reprinted here for those, who value content over form and, hence, should feel comfortable with them.
Our aim is to solely build on the insights that can be deduced from the structure of the hexagrams.

As the result of the change, we obtain the hexagram No. 48 \emph{Ching} \qq{The Well}.  So, if
No. 47, with the additional information about the changing lines, characterizes the present
situation, then No. 48 indicates the (possible) future.  Its meaning is left, to a large degree, to
the interpretation, but the extensive commentaries to the \emph{I Ching}, known as the \emph{Ten
  Wings}, provide additional clues.

\subsection{Social relations}
\label{sec:relat-betw-agents}

We now want to utilize the structural complexity of the \emph{I Ching} for capturing the relations between agents in our small group.
To this end, we see each hexagram as the encoding of one specific group structure involving six social agents. 
Table~\ref{tab:6-9} summarizes the connection between the lines and the agents.
Agent $i$ is described by its
state, $s_{i}$, and its ability to change, $a_{i}$.  The binary state variable is
$s_{i}=+1$ for unbroken lines, and $s_{i}=-1$ for broken lines.  The ability to change is $a_{i}=-1$ if the line in the respective place represents either a 6 or a 9, and $a_{i}=+1$ otherwise.

\begin{table}[htbp]
  \centering
  \begin{tabular}{|r|c|c|c|c|}
    \toprule
    sum &    6 & 7 & 8 & 9 \\ \midrule
    three &      & 2+2+3 & 3+3+2 &  \\
    coin  &   2+2+2 & 2+3+2& 3+2+3 & 3+3+3 \\
    toss &    & 3+2+2 & 3+3+2 &  \\ \midrule
    lines & \includegraphics{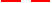} & \includegraphics{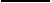} & \includegraphics{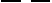} & \includegraphics{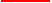} \\
    \midrule 
    probability & 0.125 & 0.375 & 0.375 & 0.125 \\ 
    \midrule
    symbol & \includegraphics{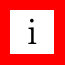} & \includegraphics{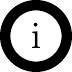} &  \includegraphics{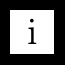} & \includegraphics{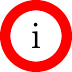} \\
    variables & $s_{i}=-1$ & $s_{i}=+1$ & $s_{i}=-1$ & $s_{i}=+1$ \\
        &    $a_{i}=-1$ & $a_{i}=+1$ & $a_{i}=+1$ &  $a_{i}=-1$ \\ \bottomrule
                                         
  \end{tabular}
  \caption{Relation between line characteristics of the \emph{I Ching} and agent variables. The symbols are used in the network diagrams of Figures~\ref{fig:GG}, \ref{fig:opinion}. }
  \label{tab:6-9}
\end{table}

The first question regards the appropriateness of matching agents of a social group with lines from a hexagram. 
Here we emphasize that the \emph{I Ching} explicitly refers to social structures when
interpreting the trigrams and the hexagrams.  For example, the 8 trigrams shown in Figure
\ref{fig:tri2} represent a family with father (a), mother (e), eldest daughter (b), middle daughter
(c), youngest daughter (d), eldest son (f), middle son (g) and youngest son (h).

Further, the 6 different places are attributed to different hierarchies of governance.  The 5th
place is the place of the ruler, the 4th place is the place of the minister.  The 3rd place has a
transitional position of limited power, but not a central one.  The 2nd place is the place of an
official far from the court, but in direct dependence from the ruler.  The 1st and the 6th place
have no specific role, although the 6th one is sometimes associated with the exalted sage, who
nevertheless plays no active role.

The second question regards information about the relation between lines.  Most effort in the
\emph{I Ching} is devoted to explain the meaning of changing lines on their respective places.
But the commentaries, notably the \emph{Ta Chuan} and Wilhelm's explanations of these commentaries, also provide some overarching insights, which we will use to define variables to characterize agents and their relations. 

\paragraph{Status. \ }
In addition to their state $s_{i}$ agents are characterized by their place (see Figure~\ref{fig:hex}), which determines their rank, or status, $r_{i}\in\{1,2,...6\}$.
We will later use the status to weight the influence of different agents.
Both agents 1 and 6 are considered of less importance, therefore we define $r_{6}=r_{1}=1$.
Higher status shall indicate more importance, $r_{1}<r_{2}<...<r_{5}$.
That means agent 5 has the highest status in the group, i.e. it has the place of the \emph{ruler}.
Additionally, agent 5 is most often the \emph{ruler of a hexagram}, which means that it defines the meaning of the hexagram, i.e. the whole group.

\paragraph{Correctness. \ }
To characterize the matching of agents to their places 
we assign a variable $c_{i}$, which is called  \emph{correctness} in the \emph{I Ching} commentary. 
$c_{i}=1$ if agents represented by weak lines are in weak places and agents represented by strong lines are in strong places.
Otherwise, if agents don't occupy correct places, $c_{i}=-1$.
Formally,
\begin{align}
  \label{eq:12}
  c_{i}= (-1)^{i+1}s_{i}
  \end{align}
This gives, as expected, $c_{i}=+1$, for ranks 1, 3, 5 if $s_{i}=+1$, and for ranks 2, 4, 6, if $s_{i}=-1$. Otherwise, $c_{i}=-1$.

The correct assignment reflects the principle that \emph{yang}
is associated with odd numbers and \emph{yin} with even numbers.
Important for us, this is considered as an \emph{equilibrium state}, which
Wilhelm explains as follows: \qq{When the firm lines are in firm places and the yielding lines
  in yielding places, a state of equilibrium exists. However, this abstract state of equilibrium
  must yield to change and reorganization when the time demands it. The time, that is, the total
  situation represented by a hexagram, plays an important role in regard to the positions of the
  individual lines.} \citep[p. 626]{WB}
We note that (i) the  equilibrium is not necessary the most stable state
and (ii) in an equilibrium state both \emph{yin} and
\emph{yang} alternate and are present in equal proportion.

\begin{figure}[htbp]
  \centering
  \includegraphics[width=0.26\textwidth]{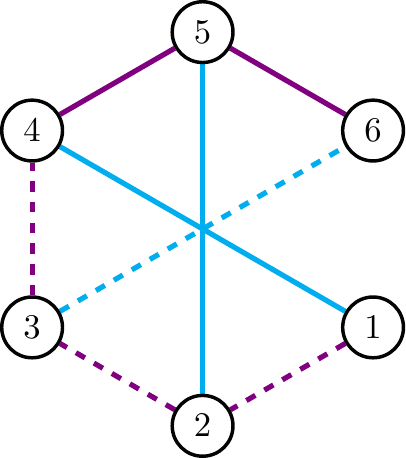}(a)
  \hspace{2cm}
  \includegraphics[width=0.26\textwidth]{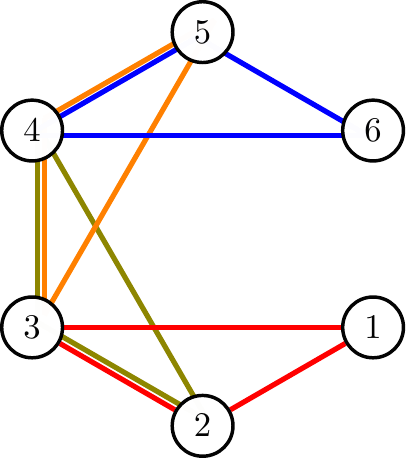}(b)
\caption{Group with (a) correspondence relations $g_{ij}$, Eqn.~\eqref{eq:14} (cyan) and  hold together relations $h_{ij}$, Eqn.~\eqref{eq:146} (purple).
    Dashed lines indicate negative relations, solid lines indicate relations that can switch between  positive/negative. (b) Triadic group structures following from the primary trigrams $T_{123}$,
    $T_{456}$ and the nuclear trigrams $T_{234}$, $T_{345}$ shown in Figure~\ref{fig:hex}.
    The networks (a) and (b) form a two-layer network. 
    }
  \label{fig:4triads}
\end{figure}

\paragraph{Correspondence. \ }
While $c_{i}$ and $r_{i}$ characterize the \emph{agent}, the \emph{I Ching}  also mentions some special \emph{relations} between agents, shown in Figure~\ref{fig:4triads}(a). 
\qq{The close relationships between the lines are those of correspondence and of holding together.
  According to whether the lines attract or repel one another, good fortune or misfortune ensues, in
  all the gradations possible in each case}. \citep[p. 666]{WB}.
Correspondence, denoted by $g_{ij}$, occurs \emph{between triads}, i.e. between agents of the lower and the upper trigram, provided they have different $s_{i}$.
\qq{As a rule, firm lines correspond
with yielding lines only, and vice versa. The
following lines, provided that they differ in kind,
correspond: the first and the fourth, the second and
the fifth, the third and the top line. Of these, the
most important are the two central lines in the second and the fifth place, which stand in the
correct relationship of official to ruler, son to father,
wife to husband.} \citep[p. 677]{WB}
Figure~\ref{fig:4triads}(a) shows the respective network. 

Whether correspondence relations are seen as favorable or not results to some extent already from the  equilibrium principle, with a noticeable exception.  
The relation between agents 2 and 5 can be favorable even if agent 5 is represented by a weak line and agent 2 by a strong line, because 
this  can be interpreted as \qq{support}.
A correspondence relation between a strong line in the
1st place and a weak line in the 4th place is also considered favorable, while the same relation between a weak line in the 1st place and a strong line in the 4th place is not favorable, because these agents occupy the wrong places, according to  the equilibrium condition. 
Correspondence relations between agents 3 and 6 are considered rare and not favorable.

To formalize these relations, we specify the resulting signs $g_{ij}$ as follows:
\begin{align}
  \label{eq:14}
  g_{25}&=\, + \big\{ 1-\Theta[s_{2}s_{5}] \big\}\nonumber \\
  g_{36}&=\,  - \big\{1-\Theta[s_{3}s_{6}] \big\} \\
  g_{14}&= c_{4} \big\{1-\Theta[s_{1}s_{4}] \big\} \nonumber 
\end{align}
 $\Theta[x]$ is the Heaviside
function which returns 1 if $x > 0$ and 0 if $x\leq 0$.
$1-\Theta[s_{2}s_{5}]=0$ because the principle of correspondence would be violated if both $s_{2}$ and $s_{5}$ have the same sign. 
If they have different signs, it does not really matter whether agents 2 and 5 are at their correct places, therefore we do not have a dependence on $c_{i}$.

For 
$g_{36}$ follows a similar argument, but it should always be negative.
$g_{14}$ can be positive or negative dependent on whether the agents occupy their correct places.
This can be indicated by either $c_{4}=\pm 1$ or $c_{1}=\pm 1$ because a configuration with $c_{4}\neq c_{1}$ implies $s_{4}=s_{1}$ and is therefore excluded.

\paragraph{Holding together. \ }
A second set of relations $h_{ij}$ involves neighboring agents \emph{within triads}, also shown in 
Figure~\ref{fig:4triads}(a).
Again, it is required that neighboring agents have opposite $s_{i}$. 
Favorable relations occur between a weak line in 4th place and a
strong line in 5th place, which again matches the equilibrium condition.
The opposite, a strong line in the 4th place and a weak line in the 5th place, is considered less favorable in many cases, but exceptions exist.

Additionally, holding together is also favorable between agents 5 and 6, provided that agent 5 is represented by a weak line and agent 6 by a strong line.
The opposite relation is considered negative.
For the lower triad, a holding together between agents 1 and 2, 2 and 3, 3 and 4 is  considered as rare and never positive. 

To formalize these relations, we specify another set of signs $h_{ij}$ as follows:
\begin{align}
  \label{eq:146}
  h_{45}&=+c_{4} \big\{ 1-\Theta[s_{4}s_{5}] \big\}\nonumber \\
  h_{56}&= - c_{6} \big\{1-\Theta[s_{5}s_{6}] \big\} \nonumber \\
  h_{12}&= \;\;\;\,  - \big\{1-\Theta[s_{1}s_{2}] \big\} \\
  h_{23}&= \;\;\;\,  - \big\{1-\Theta[s_{2}s_{3}] \big\} \nonumber \\
  h_{34}&=\;\;\;\,  - \big\{1-\Theta[s_{3}s_{4}] \big\} \nonumber 
\end{align}

It should be noted that the two sets of relations are only partially aligned to the equilibrium condition. 
For instance, a weak line in 4th place would increase both $h_{45}$ and 
$g_{14}$.
On the other hand, a strong line in 5th place would result in a negative $h_{56}$. 
It is precisely this tension between different relations which now makes the question about the  stability of the group much more interesting.

\section{Evaluation of group structures}
\label{sec:eval-group-struct}

\subsection{Social impact}
\label{sec:agent-profiles}

\begin{figure}[htbp]
  \begin{center}
    
    \includegraphics[width=0.26\textwidth]{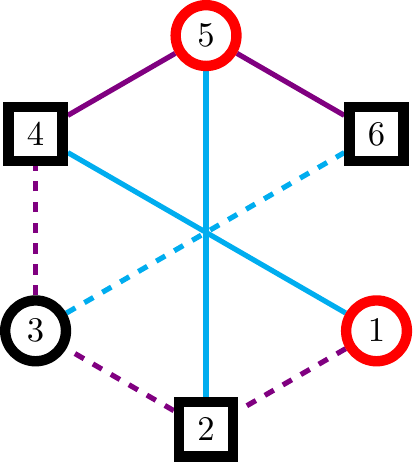}(a)\hspace*{-0.3cm} \raisebox{2cm}{\fontsize{45}{5}{$\Rightarrow$}}
    \includegraphics[width=0.26\textwidth]{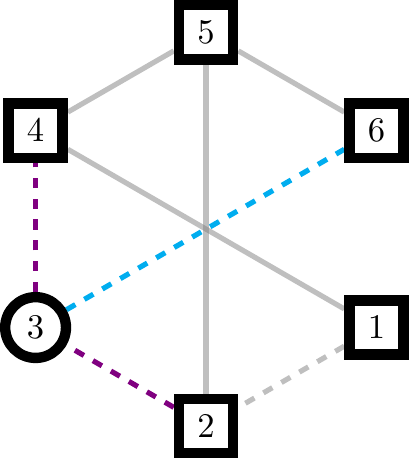}(b)

    \bigskip

    \includegraphics[width=0.26\textwidth]{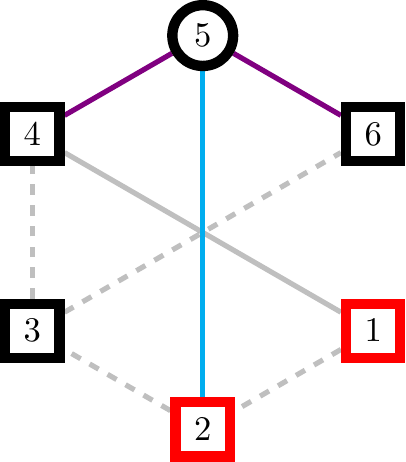}(c)\hspace*{-0.3cm} \raisebox{2cm}{\fontsize{45}{5}{$\Rightarrow$}}
    \includegraphics[width=0.26\textwidth]{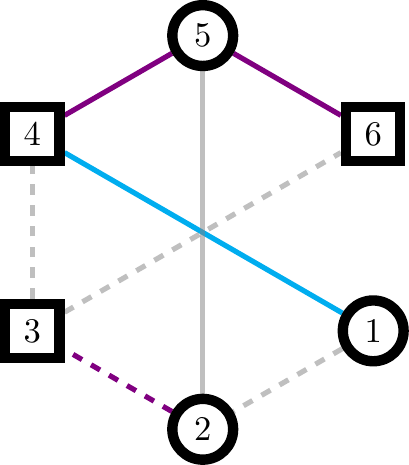}(d)

  \end{center}
  \caption{Group configurations at times $t_{0}$ (a,c) and $t_{1}$ (b,d). \scalebox{2}{$\Box$}
    indicate agents with $s_{i}=-1$, \scalebox{1.2}{$\bigcirc$} agents with $s_{i}=+1$.  Red borders
    indicate agents with $a_{i}=-1$ (ability to change), black borders agents with $a_{i}=+1$ (no
    change). Comparing these plots with the network of social relations, Figure~\ref{fig:4triads}(a), we see that some of the $g_{ij}$, $h_{ij}$ (colored in gray) are equal to zero, depending on the configuration.}
  \label{fig:GG}
\end{figure}

The group \emph{relations}  $g_{ij}$, $h_{ij}$ shown in Figure~\ref{fig:4triads}(a) are now specified. 
To have some illustrative examples for group \emph{configurations} Figure~\ref{fig:GG} depicts two slightly different configurations at time $t_{0}$.
Because some
agents have the ability to change their state, i.e. $a_{i}=-1$, we obtain different configurations at
time $t_{1}$, which are also shown in Figure~\ref{fig:GG}.
We can still not estimate the stability of these groups because, from our perspective,  the \emph{signs} of the relations are not sufficient for this.
Instead, relations should depend on the respective agents, specifically on their state, their correctness, their status.
Further, all agents have more than one relation and therefore are subject to different influences at the same time.
To decompose the network of relations shown in  Figure~\ref{fig:GG} is therefore not appropriate.
We need to find ways to aggregate these different influences on the level of agents and to also include the diverse agent features.

To this end we resort to social impact theory \citep{Bushman2007,fink-96,Jackson1987} which quantifies the social impact $I_{i}$ as the difference between positive and negative \qq{forces} impacting agent $i$ from its neighbors \citep{holyst2001,lewenst-nowak-latane-92,nowak-szam-latane-90}.
Taking the example of agent 5, it receives influences from agent 4 via $h_{45}$, from agent 6 via $h_{56}$ and from agent 2 via $g_{25}$.
These influences have to be multiplied with the weights expressing the importance of the counter parties.
Hence the impact on each agent can be calculated as follows:
\begin{align}
  \label{eq:21}
  I_{i}=\sum_{j=1}^{6}\big(g_{ij}r_{j} + h_{ij}r_{j}\big)
\end{align}
To operationalize $g_{ij}$, $h_{ij}$ we need to 
express the Heaviside functions.
We verify that
\begin{align}
  \label{eq:23}
  \Theta[s_{i}s_{j}] & = \sign(s_{i})\cdot \sign(s_{j})\cdot \abs{s_{i}+s_{j}}/2 
\end{align}
Thus, if $s_{i}=s_{j}$, then $\Theta[s_{i}s_{j}]=1$ and the rhs gives $+1$ as well.
If $s_{i}\neq s_{j}$, then $\Theta[s_{i}s_{j}]=0$ and the rhs gives $(-1)\times 0$. 

Whether the $I_{i}$ are positive or negative depends on the particular configuration.
The influence of state $s_{i}$ and correctness $c_{i}$ is implicitly considered already in the $g_{ij}$, $h_{ij}$. 
In addition to the status of the counter parties, $r_{j}$, 
the impact further depends on the status $r_{i}$ of agent $i$ itself, which  is called \qq{self-support}, or \qq{self-confidence} in social impact theory.
I.e. the influence of a single agent is given by $q_{i}=r_{i}+I_{i}$.
Table~\ref{tab:qi} lists  the respective values of $q_{i}$ for the different configurations discussed so far.
The group structure is conveniently summarized by means of  the respective hexagrams. 

\begin{table}[htbp]
  \centering
  \begin{tabular}{|r|c|c|c|c|c|c|c|c|}
    \toprule
  & $\;\,t\;\,$ & $\;\,q_{1}\;\,$ & $\;\,q_{2}\;\,$ & $\;\,q_{3}\;\,$ & $\;\,q_{4}\;\,$ & $\;\,q_{5}\;\,$ & $\;\,q_{6}\;\,$ \\ \midrule
(2) \includegraphics{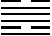}    & $t_{0}$ & $-1$ & $+3$ & $0$ & $+4$ & $+6$ & $-7$     \\
(4a) \includegraphics{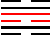} & $t_{0}$ & $-5$ & $-2$ & $-3$ & $-3$ & $+4$ & $-4$    \\
    (4b) \includegraphics{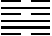}   & $t_{1}$ & $-1$ & $+1$ & $-2$ & $+6$ & $+8$ & $-7$    \\
    (6a) \includegraphics{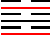} & $t_{0}$ & $+3$ & $+3$ & $-4$ & $+10$ & $+10$ & $-7$  \\
(6b) \includegraphics{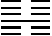}   & $t_{1}$ & $+1$ & $-1$ & $-4$ & $+1$ & $+5$ & $-2$    \\
(6c) \includegraphics{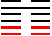} & $t_{0}$ & $+1$ & $+7$ & $+3$ & $+9$ & $+10$ & $-4$   \\
(6d) \includegraphics{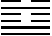}   & $t_{1}$ & $+5$ & $-1$ & $+1$  & $+13$ & $+8$ & $-4$  \\
\bottomrule
  \end{tabular}
  \caption{Influence $q_{i}=r_{i}+I_{i}$, Eqn.~\eqref{eq:21} of each agents in the respective group configuration.}
  \label{tab:qi}
\end{table}

The $q_{i}$ have a clear interpretation.
Without any social impact, the importance of each agent is reflected in its status $r_{i}$.
But group relations increase or decrease this influence.
In the favorable case, i.e. with positive social impact, the agent is supported and therefore its importance is increased.
Negative social impact, on the other hand, indicates that unfavorable relations dominate and therefore weaken the influence of the agent.
This information needs to be taken into account if we want to evaluate the stability of the group structure.

We note that group configurations in which all agents have the same $s_{i}$, either $+1$ or $-1$, do not generate any social impact.
The reason is in the definition of relations in the \emph{I Ching}, discussed in Section~\ref{sec:relat-betw-agents}, which requires agents to have opposite $s_{i}$. 
Hence in such cases the influence of an agent is simply given by its status, $r_{i}$.

 \subsection{Weighted balance condition}
\label{sec:weight-balance-cond}

So far, we have achieved to aggregate the social impact on each agent, to determine its influence, and to specify the nature of relations between agents.
But we remind that information about agents, i.e. about lines, is not sufficient to explain a hexagram.
Instead, 
we need to consider that the group structure is determined
by the 4 different \emph{trigrams} shown in Figure~\ref{fig:hex}.  One could rightly argue that these
trigrams should be represented as \emph{open} triads in a network, simply because the lower and the
upper line are not directly connected.  We nevertheless decide to represent these trigrams as
\emph{closed} triads, because the trigrams shown in Figure~\ref{fig:tri2} are seen as the
\emph{constituting units} of each hexagram.
Trigrams are building blocks, not open structures, and
their meaning comes from their connectedness and relatedness.

The network representing our group is now composed of \emph{four overlapping triads}, shown in Figure~\ref{fig:4triads}(b).
We note that these triads only contain the correspondence relations shown in Figure~\ref{fig:4triads}(a), but not the hold together relations.
Hence, triads cannot be reduced to relations, or the other way round. 
The two networks shown in Figure~\ref{fig:4triads} form a multi-layer network, which has the same agents in both layers, but the links in each layer describe different types of influences.
On the \qq{lower} layer agents are impacted by other agents via social \emph{relations}, which are the $g_{ij}$, $h_{ij}$ in our case.
This layer determines their influence values $q_{i}$, which in turn feed back to the \qq{upper} layer by determining the stability of the \emph{triads}, which can now be estimated.

Building on the calculated influence $q_{i}$ of each agent, we first introduce the \emph{weight} of a triad as a new variable:
\begin{align}
  \label{eq:24}
\Omega_{ijk} =   \big(2\Theta[q_{i}]-1\big)\ \big(2\Theta[q_{j}]-1\big)\ \big(2\Theta[q_{k}]-1\big) \ \big(\abs{q_{i}}\abs{q_{j}}\abs{q_{k}}\big)^{1/3} 
\end{align}
We note that $\big(2\Theta[q_{i}]-1\big)$ is equal to $\sign(q_{i})$ which is needed for the operationalization.

$\Omega_{ijk}$ is the \emph{signed geometric mean} of the three $q_{i}$.
It is used in  \emph{weighted balance theory} \citep{Schweighofer_2020} to consider, in addition to the signs, also the magnitude of agents' influences.
We emphasize that $\Omega_{ijk}$ results from the properties of \emph{agents}, not of their relations, i.e. it aggregates in one value the extent to which agents receive support in the whole group.
An agent with a negative $q_{i}$ is mainly determined by the negative influences from the group.
Thus, it is reasonable to assume that this agent does not actively contribute to the stability of a triad, on the contrary.
With 
two agents with negative $q_{i}$, on the other hand, it becomes obvious that the agent with positive $q_{i}$ rules the triad.
This would not hamper the stability. 
As the commentary to the \emph{I Ching} states it: \qq{
Where one alone rules, unity is present, whereas
when one person must serve two masters, nothing
good can come of it.
}\citep[p. 643]{WB}

As seen in Table~\ref{tab:qi}, the $q_{i}$ can become zero because of the integers used for the $r_{i}$.
To correct for this artifact, we set $q=0$ to the small nonzero, but positive value 0.1.
This value is chosen arbitrarily to ensure that $q=0$ would neither nullify the weight of the triad, nor change its  stability. 

Structural balance theory has calculated the stability of triads $T_{ijk}$ using simply the product of the signs of the weights $w_{ij}$, Eqn.~\eqref{eq:9}.
We have already added the weight $\Omega_{ijk}$ reflecting the influence of agents, but 
it remains to specify the respective weights $w_{ij}$ for our four triads.
From Figure~\ref{fig:4triads}(a,b) we verify that the $g_{ij}$ do not play a role in the respective triads, but the $h_{ij}$ do. 
Thus we can use information from  $h_{ij}$ to determine some of the $w_{ij}$, but we have to take into account that these $h_{ij}$ become \emph{zero} whenever the involved agents have the same $s_{i}$. 
The expressions for $w_{ij}$ have to be corrected for such cases, which is achieved by defining: 
\begin{align}
  \label{eq:22}
  w_{ij}= \big[\delta_{0,h_{ij}} + h_{ij}\big] \end{align}
$\delta_{1,s_{i}}$ is the Kronecker delta for discrete variables, $\delta_{x,y}=1$ if $x=y$. 
Therefore 
$w_{ij}=h_{ij}$ if $h_{ij}\neq 0$ and $w_{ij}=+1$ if $h_{ij}=0$.
This can be operationalized as follows: 
\begin{align}
  \label{eq:73}
  \delta_{0,y_{ij}} + h_{ij} & = 1-\sign(\abs{h_{ij}}) + h_{ij}  
\end{align}
This makes use of the fact that $\sign(0)=0$.
I.e. if $h_{ij}=0$, we arrive at $+1$, if $h_{ij}\neq 0$, we arrive at $h_{ij}$ because
$\sign(\abs{h_{ij}})=1$ by definition. 

The definition for $w_{ij}$, Eqn.~\eqref{eq:22}, sets the values $w_{13}$, $w_{24}$, $w_{35}$, $w_{46}$ for which we have no additional information, to $w_{ij}=+1$.
This deserves some discussion.
Going back to the structural stability of triads shown in Figure~\ref{fig:triads}, we note that $w_{ij}=+1$ \emph{never changes} the stability of the triad, which is determined completely by the number of negative relations.
Hence, choosing instead $w_{ij}=-1$ would definitely change the stability.  
We have no reasons for such a far reaching assumption, therefore we choose  $w_{ij}=+1$ if no other information about the relation is available.
As a side remark, with this choice it makes no difference whether we consider open triads or closed triads to calculate the stability of triads.

With this, we can define the stability of a triad as follows:
\begin{align}
  \label{eq:19}
  T_{ijk}= w_{ij}\, w_{ik}\, w_{jk} \, \Omega_{ijk}
  \end{align}
This definition is different from the one used structural balance theory, in important aspects. 
First, the relations $w_{ij}$ between agents depend on the states $s_{i}$ of the agents themselves.
Secondly, the weight $\Omega_{ijk}$ of a triad reflects the mutual impact of agents, not just from the triad, but from the whole group.   
Agents that are not \qq{balanced} themselves, i.e. have no positive $q_{i}$, cannot constitute balanced triads.
In other words, we assume that the balance of triads is grounded in agent properties, which then may impact relations, rather than postulating that relations exist independent of agent properties.

The respective values of $T_{ijk}$ for the group configurations discussed so far are shown in Table~\ref{tab:t123}. 
To estimate the stability of the whole \emph{group} from the stability of the \emph{four triads}, shown in Figure~\ref{fig:4triads}(b), we calculate the  total balance from the superposition of the constituting triads as follows:
\begin{align}
  \label{eq:17}
  T_{1-6}=\frac{1}{4}\big[T_{123}+T_{456}+T_{234}+T_{345}\big]
\end{align}
\begin{table}[htbp]
  \centering
  \begin{tabular}{|r|c|c|c|c|c|c|c|c|}
    \toprule
& $t$  & $T_{123}$ & $T_{234}$ & $T_{345}$ & $T_{456}$ & $T_{1-6}$ \\ \midrule
    (2) \includegraphics{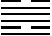}    & $t_{0}$ & $-0.67$ & $-1.06$  & $+1.34$ & $+5.52$ & $+1.28$  \\
    (4a) \includegraphics{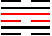} & $t_{0}$ & $-3.11$ & $-2.62$  & $-3.30$ & $-3.63$ & $-3.17$  \\
    (4b) \includegraphics{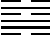}   & $t_{1}$ & $-1.26$ & $+2.29$  & $+4.58$ & $+6.95$ & $+3.14$  \\
        (6a) \includegraphics{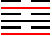} & $t_{0}$ & $-3.30$ & $-4.93$  & $+7.37$ & $+8.88$ & $+2.00$  \\
    (6b) \includegraphics{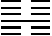}  & $t_{1}$ & $-1.59$ & $+1.59$  & $+2.71$ & $-2.15$ & $+0.14$  \\
    (6c) \includegraphics{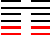} & $t_{0}$ & $+2.76$ & $+5.74$  & $+6.46$ & $+7.11$ & $+5.52$  \\
    (6d) \includegraphics{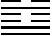}  & $t_{1}$ & $+1.71$ & $+2.35$  & $+4.70$ & $+7.47$ & $+4.06$  \\
    \bottomrule
  \end{tabular}
  \caption{Weighted stability of triads and groups discussed so far.
    \label{tab:t123}
}
\end{table}

If we compare the stability of the different group configurations, we find stability in most cases, despite the fact that sometimes only half of the four triads are stable themselves.
This results from the dominating influence of agent 5, which is the \qq{ruler} and has the highest status $r_{i}$.
Hence, the triad $T_{456}$, where agent 5 has the central position,  usually had the highest values in our examples and impacts the group stability the most.
This is a consequence of the social structure underlying the hexagrams of the \emph{I Ching}.
A good governance, in particular a good ruler, plays a major role for obtaining stable structures.

To better understand the values of the $T_{ijk}$, let us take a look at some special cases.
What would be the result if \emph{all} lines are either weak or strong, i.e. all agents have either $s_{i}=+1$ or $s_{i}=-1$?
According to our above discussion, these should be very stable configurations.
Indeed, we find that all triads have positive values for $T_{ijk}$, with an average $T_{1-6}=2.83$, regardless of their $s_{i}$.
Let us consider now a configuration where the upper trigram has only weak lines, i.e. $s_{i}=-1$ for agents 4, 5, 6 and the lower trigram has only strong lines, i.e. $s_{i}=+1$ for agents 1, 2, 3.
Then, we find that $T_{123}=T_{456}=-4.12$, while $T_{234}=T_{345}=+4.12$, which means $T_{1-6}=0$.
That means, the stability of a \emph{trigram} that contains only weak or only strong lines, would be considered \emph{negative}, but the stability of a \emph{hexagram} that contains only weak or strong lines would be \emph{positive}.

Now, we invert this configuration, i.e. the upper trigram is made up by only strong lines and the lower one by only weak lines.
In this case we find that $T_{123}=T_{456}=+3.48$, while $T_{234}=T_{345}=-3.48$ and $T_{1-6}=0$.
Now, both the upper and the lower trigram are considered stable, while the nuclear trigrams are considered unstable. 
Compared to the previous case, this time the strong place 5 is occupied by a strong line, which changes the character of the whole hexagram.
This points again to the fact that the hexagram is not simply the superposition of the upper and lower trigrams.

Eventually, we take a look at the configuration of the equilibrium condition, which is given in the the fourth row of Table~\ref{tab:t123}.
All agents are at their correct places.
But the two lower triads are unstable because there is no positive relation between the involved agents.
Still, the stability of the whole group is positive.
Now, we invert the equilibrium condition and put all agents at the wrong places.
What would be the result?
We find that now \emph{all} triads become stable, and the resulting stability $T_{1-6}=4.53$ is much higher than for the equilibrium case. 
This seems to be quite counter intuitive.
But in the anti-equilibrium case in the lower and the upper triad two agents with a negative $q_{i}$ appear.
According to the definition of $T_{ijk}$, this is not considered a negative constellation, hence all triads have a positive stability.

Is this in line with the \emph{I Ching?}
The two hexagrams are \includegraphics{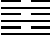}, No. 63, \qq{After Completion}, which refers to the equilibrium configuration, and  \includegraphics{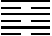}, No. 64, \qq{Before Completion}, which is the anti-equilibrium.
\qq{After Completion} is considered less favorable, because any change would destroy the equilibrium.
As the commentary by Wilhelm states it: 
\qq{It is a time of very great cultural
development and refinement. But when no further
progress is possible, disorder necessarily arises,
because the way cannot go on.}\citep[p. 1268]{WB}
\qq{Before Completion}, on the other hand, is considered as positive because it has the favorable change still ahead. 
\qq{Outwardly viewed, none of the lines appears in its
proper place; but they are all in relationship to one
another, and order stands preformed within, despite
the outward appearance of complete disorder.}\citep[p. 1276]{WB}
For this reason, No. 64 is the last hexagram of the \emph{I Ching}, according to the \emph{King Wen} order, and it presents an optimistic outlook.

\section{Resilience of small groups}
\label{sec:resil-small-groups}

\subsection{Quantifying resilience}
\label{sec:quant-resil}

Using the above examples of group configurations summarized in Table~\ref{tab:t123}, we now aim to quantify the resilience of small groups.
In general, resilience describes the ability of a system to maintain and possibly even increase its robustness when facing a change.
We propose that a resilience measure should be composed of a structural
component that captures the \emph{robustness}, $R$, and a dynamic component that captures the
\emph{adaptivity}, $A$, of a system.
This raises the question about suitable proxies for our groups. 

To proxy robustness, we use our weighted stability measure $T_{1-6}$, Eqn.~\eqref{eq:17}, which
measures balanced triads, considering the positive or negative social impact of agents.
The robustness measure $R$ should be always positive and scaled to values between 0 and 1, for comparison.
To obtain this, we scale the positive and negative values of our stability measure $T_{1-6}$ as follows: 
\begin{align}
  \label{eq:10}
R(T_{1-6})=\frac{1}{1+e^{-\beta T_{1-6}}}
\end{align}
This results into $R \to 0$ for very negative $T_{1-6}$ and $R\to 1$ for very positive $T_{1-6}$, while $T_{1-6}=0$ returns $R=0.5$.  
The parameter $\beta$ allows to adjust the slope, i.e. the sensitivity of robustness against changes of $T_{1-6}$.
We use $\beta=0.2$, larger values of $\beta$ lead to a step-like dependence.

Adaptivity $A$ is measured in our model by fraction of agents that are able to change their state, i.e. by the number of $a_{i}=-1$.
\begin{align}
  \label{eq:7}
  A(t_{0})=1-\frac{1}{n}\sum_{i}\Theta[a_{i}(t_{0})]
  \end{align}
The sum counts all agents that are not able to change their state.
It should be noted that adaptivity is a property of the group at time $t_{0}$ because
we have no information about the $a_{i}$ at a later time.

How do robustness and adaptivity impact resilience?
If all agents only have $a_{i}=+1$, the group is not adaptive, i.e. it cannot change.  This does not imply  stability,  because robustness can still be low.
If a
group is neither adaptive nor robust, it has a large chance to simply collapse.
On the other hand, the fact that a group is adaptive does not imply that it will increase its robustness.
As the examples in this paper have shown, configurations at $t_{1}$ can also be worse with respect to
their stability.
Conversely, if a group is not adaptive, this may not be bad as long as the group
is sufficiently robust. To conclude, adaptivity bears the chance to improve the robustness of the
group, but also the risk to reduce it.

Resilience, as a quantitative measure, should try to balance the influence of both robustness and adaptivity.
It should be low if robustness and adaptivity are low, because the chances to improve the situation for the
group are low in such cases.
It should be also low if robustness and adaptivity are high, because the risk
to destroy a good situation for the group is high in these cases.
A group with low
robustness has nothing to lose, thus a high adaptivity can only improve the situation.
A
group with high robustness has a lot to loose, thus adaptivity should be low to keep resilience
high.  These considerations determine us to quantify resilience as \begin{align}
  \label{eq:5}
  \mathcal{{R}}(A,R)=R (1-A)+A(1-R) 
  \end{align}
With the normalized $R$ and $A$, this gives resilience values between 0 and 1.
As requested, $\mathcal{R}$ is high if either $R$ is high and $A$ is low, or if $A$ is high and $R$ is low or if both have intermediate values.  
Table~\ref{tab:ra}  presents the robustness, adaptivity and resilience values obtained for the sample group configurations.
We remind that for time $t_{1}$, i.e. for future configurations, we do not have
information about adaptivity, hence we cannot compute their resilience.
But we can compare current and future  configurations with respect to their robustness.
\begin{table}[htbp]
  \centering
  \begin{tabular}{|r|c|c|c|c|c|c|}
    \toprule
 & $t$ & $R$ & $A$ & $\mathcal{{R}}$ \\ \midrule
    (2) \includegraphics{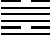}    & $t_{0}$ & $0.57$ & $--$    & $--$    \\
    (4a) \includegraphics{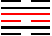} & $t_{0}$ & $0.35$ & $0.33$  & $0.45$  \\
    (4b) \includegraphics{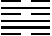}   & $t_{1}$ & $0.65$ & $--$    & $--$    \\
    (6a) \includegraphics{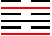} & $t_{0}$ & $0.60$ & $0.33$  & $0.53$   \\
    (6b) \includegraphics{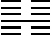}   & $t_{1}$ & $0.51$ & $--$    & $--$     \\
    (6c) \includegraphics{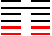} & $t_{0}$ & $0.75$ & $0.33$  & $0.58$  \\
    (6d) \includegraphics{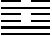}   & $t_{1}$ & $0.69$ & $--$    & $--$    \\
    \bottomrule
  \end{tabular}
  \caption{Robustness $R$, Eqn.~\eqref{eq:10}, adaptivity $A$, Eqn.~\eqref{eq:7}, and resilience $\mathcal{R}$, Eqn.~\eqref{eq:5}, of the groups discussed so far.
    \label{tab:ra}
}
\end{table}

These numbers shall illustrate that we are able to \emph{compare} initial group configurations.
The absolute numbers depend on our operationalization. 
We could choose additional parameters to weight the influence of robustness and adaptivity, we could also operationalize $A$ and $R$ in different ways, if there is evidence for it.  

It should be noted that in the example (4a) an increase of adaptation will always lead to an improved resilience, simply because the robustness is low.
This is in line with the requirements we have set up for the resilience function.
In the example (6c), on the other hand, an increase of adaptation will always lead to a decreased resilience, because robustness is high.
This raises a final question, namely how likely is it to observe an increase or decrease of resilience, which will be tackled in the following.

\subsection{Stochastic approach}
\label{sec:stochastic-approach}

To calculate the probability of finding certain group structures, as expressed by their hexagrams, 
let $p_{i}=p(s_{i}=+1,t_{0})$ be the probability to find agent $i$ in state $s_{i}=+1$ initially.
$i$ refers again to the position of the agent.
Going back to Table~\ref{tab:6-9} that summarizes how lines are assigned to places in the \emph{I Ching}, we verify that $p(s_{i}=+1,t_{0})=0.5$ for all agents.
I.e. there is no bias towards either $+1$ or $-1$.
When building a group structure of 6 agents, the $p_{i}$ are determined independently, therefore the probability for any group configuration $\mathbf{S}(t)$ at $t_{0}$
is:
\begin{align}
  \label{eq:25}
  p(\mathbf{S},t_{0}) & =
  p(s_{1}=+1,...,s_{6}=+1,t_{0})  \nonumber \\
  & = \prod_{i=1}^{6}p(s_{i}=+1,t_{0}) =\left(\frac{1}{2}\right)^{6}=\frac{1}{64}=0.015
\end{align}
Hence, every of the 64 possible group configurations has the same low probability of 1.5\% to appear.
The more interesting question, from the perspective of the \emph{I Ching}, regards the probability that a hexagram can also \emph{change}.
In our terms this is denoted as \emph{adaptivity} and means that the group structure is \emph{resilient}.
From Table~\ref{tab:6-9} we see that a changing line appears only with a probability $p=p(a)=0.25$.
This leads to the probability distribution for adaptivity, $P(A)=P(n,k)$, where $n$ is the number of agents and $k$ is the number of agents that can change their state: , i.e. $k/n=A$:
\begin{align}
  \label{eq:266}
 \sum_{k=0}^{n} P(n,k)  = &\sum_{k=0}^{6} \binom{6}{k} p^{k} (1-p)^{6-k}  = 
\left(\frac{1}{4}\right)^{0}\left(\frac{3}{4}\right)^{6} +                                                          6\left(\frac{1}{4}\right)^{1}\left(\frac{3}{4}\right)^{5}
    +15 \left(\frac{1}{4}\right)^{2}\left(\frac{3}{4}\right)^{4}+ \nonumber \\
 &  +20 \left(\frac{1}{4}\right)^{3}\left(\frac{3}{4}\right)^{3}+ 
  15 \left(\frac{1}{4}\right)^{4}\left(\frac{3}{4}\right)^{2}
    +6 \left(\frac{1}{4}\right)^{5}\left(\frac{3}{4}\right)^{1} +
   \left(\frac{1}{4}\right)^{6}\left(\frac{3}{4}\right)^{0} \\
= & \  0.178 + 0.356 + 0.296 + 0.131 + 0.032 + 0.004 + 0.0002  =  1 \nonumber
\end{align}
It is worth to write out the binomial distribution explicitly, to verify that the chances to find a group configuration with, e.g., two agents that can change their state (indicated in red in our examples) are quite  low, precisely $0.019\times 15$, where 15 refers to the different possibilities to pick 2 out of 6 agents.
If adaptivity is expressed by the number of possible changes, then the probability to find a larger adaptivity decreases quickly. 
The good news is that only in 17.8\% of all cases, we should \emph{not} expect any change.
The majority of all possible group configurations has the ability to change, i.e. it can be expected to be resilient according to our definition. 

To discuss a decrease or increase of robustness would require us to consider (i) a \emph{specific} group configuration and (ii) the probability that \emph{specific} agents change their state.
We illustrate the stochastic dynamics using the example shown in Figure~\ref{fig:change}, with the respective numbers given in Table~\ref{tab:ra} (4a, 4b). 
The probability of finding a specific group configuration $\mathbf{S}$ at time $t_{1}$ follows from the master equation:
\begin{align}
  \label{eq:3a}
  p(\mathbf{S},t_{1})&= p(\mathbf{S},t_{0}) \big[1-p(A)\big] +\sum\nolimits_{\mathbf{S^{\prime}}} p(\mathbf{S}^{\prime},t_{0})\ p(\mathbf{S}|\mathbf{S}^{\prime})
\end{align}
The first term on the rhs gives the probability that the configuration $\mathbf{S}$ already exists and does not change.
The second term gives the probability of other configurations $\mathbf{S}^{\prime}$ multipied by the transition probability to change from $\mathbf{S}^{\prime}$ into $\mathbf{S}$ in the next time step.  
The summation goes over all possible configurations $\mathbf{S}^{\prime}$.
From Eqn.~\eqref{eq:25} we know that the $p(\mathbf{S}^{\prime},t_{0})$ are the same, so they can be taken out of the sum. 
Using Eqn.~\eqref{eq:266} we further know that
\begin{align}
  \label{eq:2}
  1-p(A)= \binom{6}{0} \left(\frac{1}{4}\right)^{0}\left(\frac{3}{4}\right)^{6}
  \;;\quad
  \sum_{\mathbf{S}^{\prime}} p(\mathbf{S}|\mathbf{S}^{\prime}) = \sum_{k=1}^{6} \binom{6}{k}   \left(\frac{1}{4}\right)^{k}\left(\frac{3}{4}\right)^{6-k}
\end{align}
But the single transition probabilities $p(\mathbf{S}|\mathbf{S}^{\prime})$ need to be determined dependent on the specific configuration $\mathbf{S}^{\prime}$. 
To make use of Eqn.~\eqref{eq:3a}, we fix the initial condition, i.e. we specify 
$\mathbf{S}^{\prime}(t_{0})=\mathbf{\hat{S}}=\{-1,+1,-1,+1,+1,-1\}$ from the example (4a).
Thus, $p(\mathbf{S}^{\prime}=\mathbf{\hat S},t_{0})=1$ instead of 0.015 as given by Eqn.~\eqref{eq:25}.  
To obtain the configuration $\mathbf{S}(t_{1})=\{-1,+1,+1,-1,+1,-1\}$ from the example (4b), a change of agents 3 and 4 is needed, while all other agents should not change.
Because the probabilities to change are independent for all agents, we have:
\begin{align}
  \label{eq:3}
  p(\mathbf{S}|\mathbf{\hat S})=\left(\frac{1}{4}\right)^{2}\left(\frac{3}{4}\right)^{4}=0.019
\end{align}
This is the probability that the group described by the configuration $\mathbf{\hat S}$ will improve its robustness towards the configuration $\mathbf{S}$.
From Table~\ref{tab:ra}, we see that the robustness would improve from 0.35 to 0.65. 

In more general terms, instead of a fixed initial condition we have to consider (i) the probability that the configuration (4a) appears, which is $p(\mathbf{\hat S})=0.015$, and (ii) that the specific transition between the configurations (4a) and (4b) occurs, Eqn.~\eqref{eq:3}. 
This results in $0.015\times 0.019=2.8\times 10^{-4}$.
This lower bound can be compared with the probability that our specific group configuration $\mathbf{\hat S}$ has two \emph{randomly chosen} agents changing.
It would give $0.015\times 0.296=44.4\times 10^{-4}$, because there are 15 different possibilities to choose two agents, Eqn.~\eqref{eq:266}.
This could be seen as an upper bound.

This short exercise allows two insights.
First, we are able to calculate the probability to find groups with the ability to change, which is 82.2\% and quite high.
We remind that this ability is the precondition for resilience. 
Secondly, we can, for every possible group configuration $\mathbf{S}^{\prime}$, calculate its robustness $R$, as we have shown above, from calculating its stability $T_{1-6}$, which includes in a weighted manner the social impact and importance of all agents, $q_{i}$.
We can then identify all those group configurations $\mathbf{S}$ that have a larger  robustness than a given configuration $\mathbf{\hat S}$, and we can calculate the transition probabilities for a possible change from $\mathbf{\hat S}$ to a desired configuration $\mathbf{S}$.
These transition probabilities depend on the probability to have a given adaptivity $A$, in general, but also on the combinatorial probability to find the right agents for a change, Eqn.~\eqref{eq:266}. 
As we have seen, such probabilities are rather small.
But the procedure is straightforward, and it gives us for every group configuration the probabilities to improve or to decrease its robustness $R$ dependent on its adaptivity $A$. 

One may find that the probabilities for adaptivity should be modified. 
One possible option is already considered in the  \emph{I Ching}.
In addition to the three coins toss described in Table~\ref{tab:6-9}, which is a rather \qq{modern} way of generating randomness, an older procedure uses 50 yarrow sticks in a more complicated manner.
Instead of the 8 possibilities to toss 3 coins listed in Table~\ref{tab:6-9}, we then have 64 possibilities and the probabilities are distributed as follows:
$p(6)=4/64$, $p(7)=20/64$, $p(8)=28/64$, $p(9)=12/64$ \citep{wilhelm-1972-sinn-i-ging}.
This still keeps the ratio of 1:1 for broken and unbroken lines and the ratio 1:3 for the probability of change.
But it generates a bias of 3:1 towards the change of unbroken lines, which was 1:1 when using the coins.
Or, the other way round, if a \emph{yang} line is chosen, it has a 3:5 chance to change, while a \emph{yin} line has a 1:7 chance to change, which was 1:3 for both, before.

\section{Discussion}
\label{sec:disc}

To first address a misunderstanding, this paper reveals nothing new about the \emph{I Ching}, which is  around for more than 2000 years.
But in order to interpret the hexagrams of the \emph{I Ching} in terms of group structures, we had to
formalize the subtle relations between the different lines.
To further model group structures we needed additional assumptions summarized in the following.
The should be seen as examples to demonstrate how the model works, rather than sociologically founded certitudes.
The following building blocks are used to specify our agent-based model:

\paragraph{State variables:} Agents differ from dots in a network in their internal degrees of freedom. For the $s_{i}$ we used a binary variable, to express a fundamental duality.
  This is a common assumptions in many spin-like models, such as the voter model \citep{fs-voter-03} or formal models of social impact \citep{Belaza2017,holyst-kacp-fs-00,kohring96,lewenst-nowak-latane-92}. 
  Alternatively, it is possible to use multi-dimensional vectors and continuous values for these state variables, as used for instance in 
  models of multi-dimensional opinion dynamics \citep{Starnini2021,schweighofer20,Schweighofer_2020}.
  This would only complicate the analysis, but not change the model fundamentally. 

\paragraph{Dynamics:} We need assumptions of \emph{how} and \emph{when} the state variables of agents change.
    We used a binary agent variable $a_{i}$ to indicate which agents could change their $s_{i}$.
    Because the $s_{i}$ are binary variables, the direction of change was already fixed, and a deterministic dynamics was used.
    Similar to the basic voter model, stochasticity results from the random sampling of  (i) the $s_{i}$ and (ii) the $a_{i}$, to determine the initial state. 
    We provided one possible scenario for the initial sampling.
    We note that 
    in our model change takes place with a rather low probability ($p=0.25$).
This is not a drawback, it is a feature of our model which wants to study social relations. 

\paragraph{Relations:} Social relations differ from mere interactions, which are rather frequent and often random.
  Relations have to be built up over time, for instance as  the result of many interactions.
  It is assumed that they express the fundamental quality of a relationship, last longer and change less often.
  Most important, relations are not independent of the subjects of the relationship, i.e. the agents.
  In our model, we have used heuristic arguments to assign relations  which take into account (i) the internal states of both agents, (ii) their positions in the social network, (iii) their social status, (iv) their \qq{correctness} as a measure how well they fit their position.
  These assumptions can any time be replaced with better grounded ones, if further information is available.
  It just needs to specify the $w_{ij}$ in a more profound way.

  \paragraph{Heterogeneity:} Agents in our model are heterogeneous, i.e. they vary in their properties, notably in their state, in their ability to change, their social status, their network position, their preferences for relations with other agents.
  Every model of social groups has to take these features into account.
  The elements of  social systems are not atoms, but individuals.
  Our model makes some suggestions how to implement this heterogeneity without resorting to random assignments.
  Our results demonstrate that differences in agents matter for determining the social impact and the resulting group stability. 

\paragraph{Social impact:}
Agents are  involved in different relations at the same time.
Focusing only on isolated triads means to decompose a social network, which is often done to describe large networks in a mean-field approach \citep{Singh2016,papanikolaou2022consensus,battiston2020networks}.
But for small social groups, the focus of our paper, this decomposition can hardly be justified.
Therefore we need assumptions how to aggregate the effect of simultaneous relations.
To solve this problem, we used social impact theory which has the advantage of being empirically and theoretically founded \citep{fink-96,Bushman2007,latane-bourg-96,Jackson1987,Finsterbusch1982}.
The resulting influence values for each agent depend on the social relations, but also on the importance of the involved agents.

Secondly, we need assumptions of how the aggregated social impact feeds back on the stability of groups.
We see this as a feedback between two layers of a multi-layer network.
The influence values of agents which result from the first layer impact the triadic group structures on the second layer.
We calculated the stability of triads in a novel manner, using assumptions from weighted balance theory.
As a main contribution of our paper,  
this solution incorporates the heterogeneity of agents and returns a weighted measure  for the stability of groups.

\paragraph{Resilience:}
Only after specifying the dynamics of change and the stability of the group we are able to tackle one of the open questions in social science, namely the resilience of social organizations.
We follow a very general approach to describe resilience as an optimal mixture of  robustness and adaptivity.
Without the ability to adapt, systems can be stable or unstable, but they are not resilient, i.e. they cannot respond to  internal or external changes.
We have proposed a novel functional form to quantify the resilience of groups and applied it to the different group configurations used in this paper for illustrative purposes.
Further, thanks to the simple initial setup chosen for our model, we were able to calculate the probabilities for certain group configurations to increase or to lower their resilience. 

\bigskip

Our model leaves out, on purpose, one major challenge for modeling group dynamics, namely the emergence of group structures and the evolution of social networks.
Instead, we used a static group structure, to only focus on the relations between agents.
It is reasonable to assume that a \emph{change} of these group structures may occur at a different time scale and should therefore be handled separately.
There are already agent-based models to describe the initial formation of social groups, the addition of new members or the leave of established ones, the creation or deletion of social relations \citep{Chen2014,Agbanusi2018,Du2018,Gao2018,Gorski2017}.
On a longer time scale, these groups can merge \citep{nucleation21,Schweitzer2020} to form larger social networks, which then can be described by established network measures.

Instead, our main focus is to model \emph{existing} social relations between agents in groups with overlapping triads.
Triadic closure is one of the main features in social networks \citep{klimek2013triadic,bianconi2014triadic,brandenberger2019}.
It has been studied from various modeling perspectives, including ERGM \citep{block2019forms,Yap2015} and gHypE \citep{Scholtes2017} statistical models.
Their aim, however, is to \emph{infer} social relations from other network or agent features, whereas our model tries to quantify the \emph{impact} of such structures on the stability and the resilience of social groups.

Our approach can inspire other recent developments to model group interactions by means of higher order networks \citep{papanikolaou2022consensus,battiston2020networks}.
There, dyads, triads, etc., are seen as new types of nodes that can represent larger groups.
So far, in these models group structures are entirely determined by link structures, the agents as the social constituents are not considered.
To incorporate them into the model would require to estimate their impact on a higher order, or group, structure, which is precisely the aim of our paper.
We have proposed one way to quantify the heterogeneity of agents, their positive or negative social relations and their mutual social impact.

Where does the \emph{I Ching} come into play?
It considers group structures in terms of hexagrams, where lines can be seen as agents and places as their ranks.
Lines have a binary characteristics both regarding their state and their ability to change.
Most importantly, these hexagrams can be seen as the superposition of four overlapping trigrams.
That means, lines affect the whole hexagram and, at the same time, are impacted by all other lines via direct or indirect relations.
The meaning of a hexagram can therefore not be decomposed into the meaning of lines.
This resembles, in a nutshell, the problem of group relations. 
In our model agents belong to overlapping triads in a network, and we have to find ways to estimate the impact they exert on, and receive from,  other agents.
With this we can tackle the problem of group stability in a novel manner, namely by incorporating agent characteristics into structural balance. 

Our assumptions are inspired by the \emph{I Ching}, but they do not depend on it.
The formalism can readily applied using other assumptions for the required specifications summarized above.
Frankly, instead of coming up with arbitrary assumptions, it was quite tempting to formalize the ideas laid out in the \emph{I Ching} about relations, change, and fate. 
To some this might be seen as a quite exotic idea, to others not. 
After all, basic knowledge about the \emph{I Ching} should be rightly considered as part of literacy
and education.

\subsection*{Acknowledgements}
\label{sec:acknowledgements}

The author wishes to thank Georges Andres, Giona Casiraghi and Giacomo Vaccario for comments 
and Armin Schweitzer for TikZ support of the hexagrams.
This project was funded by the Swiss National Science Foundation
(SNF\_192746).

{\small \setlength{\bibsep}{1pt}

}

  \section*{Appendix: Examples for consensus relations}
\label{sec:opinion-dynamics}

\begin{figure}[htbp]
  \begin{center}
    
    \includegraphics[width=0.26\textwidth]{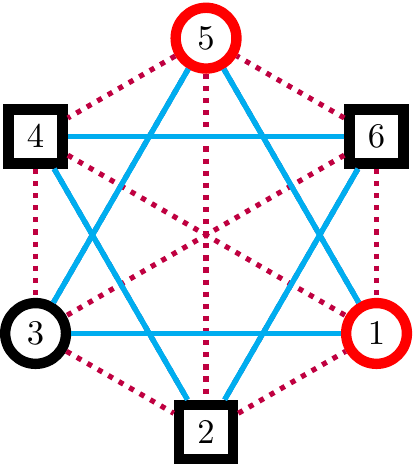}(a)\hspace*{-0.3cm} \raisebox{2cm}{\fontsize{45}{5}{$\Rightarrow$}}
    \includegraphics[width=0.26\textwidth]{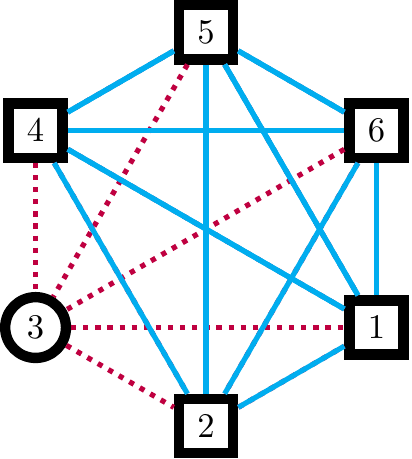}(b)

    \bigskip

    \includegraphics[width=0.26\textwidth]{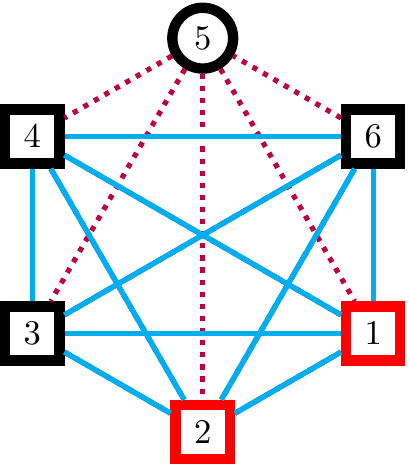}(c)\hspace*{-0.3cm} \raisebox{2cm}{\fontsize{45}{5}{$\Rightarrow$}}
    \includegraphics[width=0.26\textwidth]{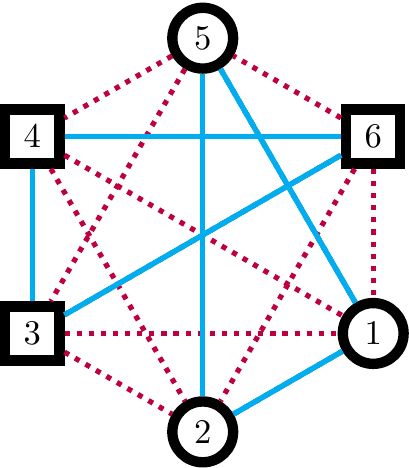}(d)

  \end{center}
  \caption{Networks following the consensus rule, Eqn.~\eqref{eq:8}, at times $t_{0}$ (a,c) and $t_{1}$ (b,d). Solid lines (cyan): consensus, dashed lines (purple): disagreement. \scalebox{2}{$\Box$}
    indicate agents with $s_{i}=-1$, \scalebox{1.2}{$\bigcirc$} agents with $s_{i}=+1$.  Red borders
    indicate agents with $a_{i}=-1$ (ability to change), black borders agents with $a_{i}=+1$ (no
    change).  }
  \label{fig:opinion}
\end{figure}

Determining the $w_{ij}$ by means of the \emph{I Ching} raises the question whether there would be simpler approaches to map agent features to relations.
As an illustrative example let us interpret the states $s_{i}$ of the agents as discrete opinions
which are assumed to have an impact on their relations.  Specifically, two agents have a positive
relation if they share the same opinion, and a negative relation if they have the opposite opinion.
The resulting $w_{ij}$ follow from:
\begin{align}
  \label{eq:8}
  w_{ij}(t)=w_{ij}\left[s_{i}(t),s_{j}(t)\right]=s_{i}(t)\ s_{j}(t)
\end{align}
Here we consider a fully connected network, as shown in Figure~\ref{fig:opinion}, i.e. all $w_{ij}=\pm1$. 
 To quantify how the change of
opinions impacts the network, we define $X(t)$ as the fraction of agents with $s_{i}=+1$ and $f(t)$ as 
the fraction of positive relations in the network:
\begin{align}
  \label{eq:4}
  X(t)=\frac{1}{n}\sum_{i} \delta_{1,s_{i}}\;;\quad f(t)=\frac{1}{m}\sum_{i<j}\Theta[w_{ij}(t)]
\end{align}
$X(t)$ is normalized to the total number of agents. 
Note that we normalize the fraction of positive relations, $f(t)$, to $m$, the number of existing links in
the network, rather than to the total number of possible links, $n(n-1)/2$.  But for the fully
connected network we have always the maximal number of possible links.   $f(t)=1$ would indicate perfect consensus,
i.e. all agents have the same opinion.  $1-f(t)$, on the other hand, indicates the fraction of
negative links coming from the dissent between agents.

The two examples of Figure~\ref{fig:opinion} use the same group configurations as in Figure~\ref{fig:GG}, but a fully connected network where the relations are determined by the consensus rule, Eqn.~\eqref{eq:8}. 
We see in (a), (b) that the fraction of positive
relations is increased at $t_{1}$, from $f(t_{0})=6/15$ to $f(t_{1})=10/15$.  But as the second
example shows, we can also arrive at the opposite.  Comparing (c) and (d), the fraction of positive
relations has decreased, from $f(t_{0})=10/15$ to $f(t_{1})=6/15$.

One can verify that in a fully connected network with 6 agents we find from an opinion fraction
$X=3/6$ for the positive relations the fraction $f=6/15$, from $X=2/6$ (or $X=4/6$) the fraction $f=7/8$ and from $X=1/6$ (or $X=5/6$) the fraction $f=10/15$.  Therefore, it depends on the configuration at
$t_{0}$, but also on the distribution of the $a_{i}$ whether an increased or decreased fraction of
positive relations is to be observed at $t_{1}$.

To quantify the topological structure with respect to balanced and unbalanced triads, we define the
fraction of balanced triads in the network as
\begin{align}
  \label{eq:6}
  F(t)=\frac{2}{(n-1)(n-2)} \frac{1}{C} \sum_{i<j<k}\Theta[T_{ijk}(t)] \,; \quad C=\frac{1}{n}\sum_{i}C_{i}=\frac{1}{n}\sum_{i} \frac{2 k_{i}}{d_{i}(d_{i}-1)}
\end{align}
where $T_{ijk}$ is given by Eqn.~\eqref{eq:9}.  
Note that we normalize the fraction of balanced triads to the number of \emph{existing} triads
rather than to the total number of possible triads, $(n-1)(n-2)/2$, in the group of size $n.$ $C$ is
the global clustering coefficient of the network, i.e. the average over the local clustering
coefficients $C_{i}$.  The latter counts the number $k_{i}$ of connected neighbors of agent $i$,
normalized by the total number of possible links in the neighborhood of $i$.  These depend on the
degree $d_{i}$ of each agent.  In a fully connected network we have $d_{i}=(n-1)$,
$k_{i}=(n-1)(n-2)/2$.  Hence $C_{i}=1$, and $C=1$.

Let us now consider the consensus rule, Eqn. \eqref{eq:8}, as above and see whether the triads in
the networks shown in Figure~\ref{fig:GG} are stable or unstable.  Obviously, only two types of
triads can result from the consensus rule, those shown in Figure~\ref{fig:triads}(a):
$w_{ij}=w_{ik}=w_{jk}=+1$, i.e. $T_{ijk}=+1$ and those shown in Figure~\ref{fig:triads}(c)
$w_{jk}=+1$, $w_{ik}=w_{ij}=-1$, i.e. again $T_{ijk}=+1$.  The other two configurations shown in
Figure~\ref{fig:triads}(b,d) cannot occur when applying the consensus rule in a fully connected network.  

Thus, we can conclude that, according to the definition of Eqn.~\eqref{eq:9}, \emph{all} 10 possible
triads in our network are \emph{stable}.  Hence, $F(t)=1$, Eqn.~\eqref{eq:6}.  This rather boring
situation can be changed only if we assume rules different from Eqn.~\eqref{eq:9} for defining the
stability of triads $T_{ijk}$.
This provides the motivation for our procedure in Section~\ref{sec:i-ching}.

\end{document}